\documentclass[twocolumn,prb,showpacs,preprintnumbers,amsmath,amssymb,floatfix]{revtex4}
\usepackage{graphicx}

\begin{document}

\title{Flatband localization in the Anderson-Falicov-Kimball model}
\author{A. M. C. Souza$^{1}$ and H. J. Herrmann$^{2,3}$}
\affiliation{$^{1}$Departamento de Fisica, Universidade Federal de
Sergipe, 49100-000 Sao Cristovao-SE, Brazil}
\affiliation{$^{2}$Departamento de F\'{i}sica, Universidade Federal
do Cear\'{a}, 60451-970 Fortaleza-CE, Brazil}
\affiliation{$^{3}$Computational Physics, IfB, ETH H\"{o}nggerberg,
HIF E 12, CH-8093 Z\"{u}rich, Switzerland}
\date{\today}

\begin{abstract}
The insulator-metal-insulator transition caused by a flatband is
analyzed within dynamical mean-field theory using the
Anderson-Falicov-Kimball model. We observe quantitative disagreement
between the present approach and previous results. The presence of
interactions enhances delocalization.

\end{abstract}


\maketitle

\section{Introduction}

Recent studies have shown a new disorder-induced insulator-metal
transition (MIT) of one-electron states, called the ''inverse
Anderson transition''.\cite{inverse,inverse2} The existence of this
transition becomes already visible for non-interacting particles
when the system has highly degenerated localized states forming a
flatband. In a flatband localized states may melt into extended
states due to disorder. For weak disorder, the localization is not a
consequence of the strength of disorder but of the flatband.
Increasing the degree of disorder localization-delocalization and
delocalization-localization transitions appear.

New aspects must appear when we also consider interactions between
particles of the system. The present work addresses this issue. The
investigation of this problem is particularly rich. The interaction
causes a Mott-Hubbard metal-insulator transition (MIT)\cite{Mott}
that competes with the Anderson transition.\cite{Anderson}

The inverse Anderson transition has been studied using various
numerical techniques, among them, level statistics,\cite{level}
$f(\alpha)$ characteristics of wave functions\cite{falfa} and the
participation radius\cite{raio} framework. However, these powerful
tools are not adequate for interacting systems in which matrices of
high order must be exactly diagonalized restricting the study to
extremely small systems.

The dynamical mean-field theory (DMFT)\cite{pap4} is a good tool to
investigate the Mott-Hubbard-Anderson MIT in lattice electrons with
local interactions and disorder. The Anderson transition has for
example been explored on the Bethe lattice considering the
Hubbard\cite{pap1} and Falikov-Kimball models.\cite{Freer} The metal
and the insulator phases are detected by analyzing directly the
local density of states (LDOS). The averaged LDOS can vanish in the
band center at a critical disorder strength for a wide variety of
averages.\cite{nos,nos2} In particular, the arithmetic mean of this
random one-particle quantity is non-critical at the Anderson
transition and hence cannot help to detect the localization
transition. By contrast, the geometric mean gives a better estimate
of the averaged value of the LDOS,\cite{pap1,pap2} as it vanishes at
a critical disorder strength and hence provides an explicit
criterion for the Anderson localization.\cite{Mon2,pap3,pap4} We
have adopted the H\"older mean and analyzed how the averaged LDOS
depends on the H\"older parameter.

In this paper, we investigate the disorder-induced insulator-metal
transition in a flatband of the Falikov-Kimball model using the
DMFT. We present the ground-state phase diagram for different values
of a parameter measuring the disorder strength and the dependence of
the MIT transition on the Coulomb repulsion between fermions.
Recently, we have applied the DMFT to this model. First, we showed
that not only the geometric mean can offer a good approximation for
the averaged LDOS providing an explicit criterion for Anderson
localization. We found that the averaged LDOS can vanish in the band
center at a critical disorder strength for a wide variety of
generalized H\"older mean.\cite{nos} Second, we have analyzed how
the presence of the next-nearest-neighbor hopping influences the
phase diagram of the ground state of this model, and, third, we
studied the main effects of the long-range correlated
disorder.\cite{nos2}

This paper is organized as follows. In the next section \ref{model}
we introduce the Anderson-Falikov-Kimball model. The DMFT approach
is described in section \ref{dmft}. In section \ref{results} we
present the results concerning the phase diagram. Finally in section
\ref{conclusions} we conclude.

\section{Model}  \label{model}
The Anderson-Falicov-Kimball model\cite{pap2} is a tight binding
model having two species of fermionic particles, mobile and
immobile, which interact with each other when both are on the same
lattice site. We introduce a local random potential for the mobile
particles, giving rise to a competition between interaction and
disorder. This model has been applied to mixed valence compounds of
rare earth, transition metal oxides, binary alloys and metal ammonia
solutions.\cite{Leu-Csa} Its Hamiltonian is
\begin{equation}
H=\sum_{i}\epsilon_{i} c_{i}^{+}c_{i} -t \sum_{(ij)} c_{i}^{+}c_{j}
+U\sum_{i}f_{i}^{+}f_{i}c_{i}^{+}c_{i},\label{Hamil}
\end{equation}
where $c_{i}^{+}$ ($c_{i}$) and $f_{i}^{+}$ ($f_{i}$) are,
respectively, the creation (annihilation) operators for the mobile
and immobile fermions at lattice site $i$. $\epsilon_{i}$ is a
random potential describing the local disorder, $t$ is the electron
transfer integral connecting nearest-neighbor sites, and $U$ is the
Coulomb repulsion when mobile and immobile particles occupy the same
site. We consider that the occupation of immobile particles is site
independent having a probability $p=1/2$. The number of mobile
particles on site $i$ is given by $n_{i}=c_{i}^{+}c_{i}$. A chemical
potential $\mu$ is introduced for the mobile subsystem to fix the
system in the half-filled band ($\overline{n_{i}}=1/2$). Here, the
energy will be given in units of the hopping element $t$ (i.e.,
$t=1$).

\section{Dynamical Mean-Field Equations} \label{dmft}

The DMFT is calculated from the Hilbert transform
\begin{equation}
G(E)=\int \frac{ d\omega N_{0}(\omega)}{\eta(E)-\omega+ 1/G(E)} \; ,
\label{g0}
\end{equation}
where $N_{0}(\epsilon)$ is the non-interacting density of states,
$G(E)$ the translationally invariant Green function, and $\eta(E)$ a
hybridization function describing the coupling of a selected lattice
site with the rest of the system.\cite{pap3} For the flatband the
non-interacting density of states is
\begin{equation}
N_{0}(E)=\frac{1}{2} [\delta (E - E_{0}) + \delta (E + E_{0})] \; ,
\label{flat}
\end{equation}
where the highly degenerated energies are $\pm E_{0}$. The relation
between $G(E)$ and $\eta(E)$ is obtained in a straightforward way
from Eq. (\ref{g0}) and Eq. (\ref{flat})
\begin{equation}
\eta(E) = \sqrt{\frac{1}{4G(E)^2} + E_{0}^2} - \frac{1}{2G(E)} \; .
\end{equation}

The LDOS is given by\cite{pap2}
\begin{equation}
P(E,\epsilon_{i})=-\frac{1}{\pi}~Im~ G(E,\epsilon_{i}) \;
,\label{eq1}
\end{equation}
where $G(E,\epsilon_{i})$ is the local $\epsilon_{i}$-dependent
Green function. For the Anderson-Falicov-Kimball model we obtain
that\cite{nos}
\begin{equation}
P(E,\epsilon_{i})=-\frac{s}{\pi} \frac{\alpha_i^2 + s^2 +
(U/2)^2}{[\alpha_i^2 + s^2 + (U/2)^2 ]^2 - U^2 \alpha_i^2} \;
,\label{eq1b}
\end{equation}
where $\alpha_{i}=E-\epsilon_{i}-r$ and $r$ and $s$ are,
respectively, the real and imaginary parts of $\eta(E)$.

We consider that $\epsilon_{i}$ is an independent random variable
characterized by a probability function
$p(\epsilon_{i})=\Phi(\Delta/2-\vert \epsilon_{i}\vert ) /\Delta$,
with $\Phi$ being the step function. The parameter $\Delta$ is a
measure for the disorder strength. The self-consistent DMFT
equations are closed inserting
\begin{equation}
G(E)= \int d\omega'\frac{P_{q}(\omega')}{E-\omega'}, \label{eq4}
\end{equation}
where
\begin{equation}
P_{q}(E)=\left\{\sum_{i} [P(E,\epsilon_{i})]^{q}\right\}^{1/q} \; .
\label{eq3}
\end{equation}
The parameter $q$ defines the $q$-H\"older average.
 The arithmetic and geometric mean are found, respectively, using
$q=1$ and $q \to 0$.

\begin{figure}
\includegraphics[width=80mm]{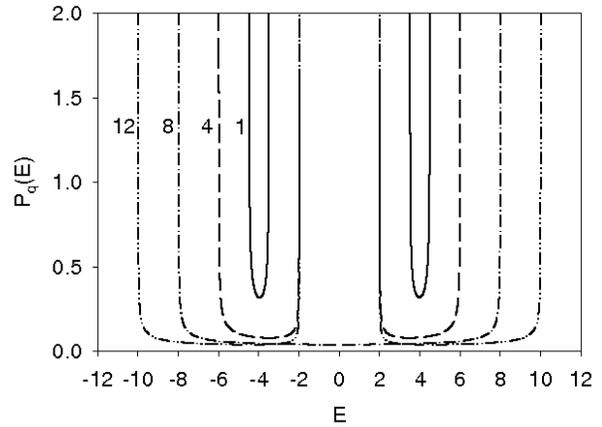}
\caption{Density of states for disorder strength $\Delta =0$ and
$U=1$, $4$, $8$ and $12$.} \label{var1}
\end{figure}

\begin{figure}
\includegraphics[width=80mm]{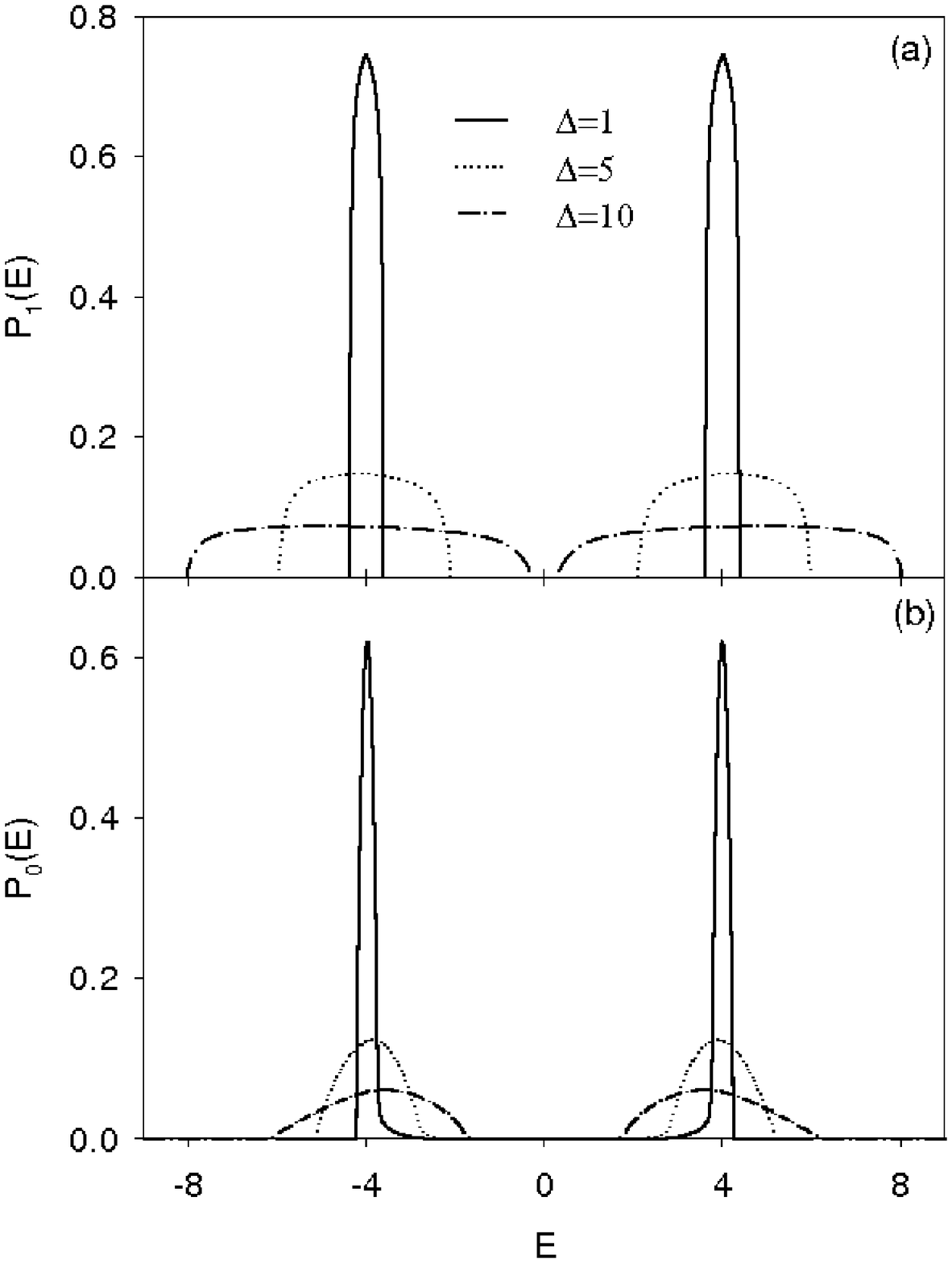}
\caption{Averaged local density of states at $U=0$ and disorder
strength $\Delta =1$, $5$, and $10$ for (a) $q=1$ and (b) $q=0$.}
\label{spec}
\end{figure}

\begin{figure}
\includegraphics[width=80mm]{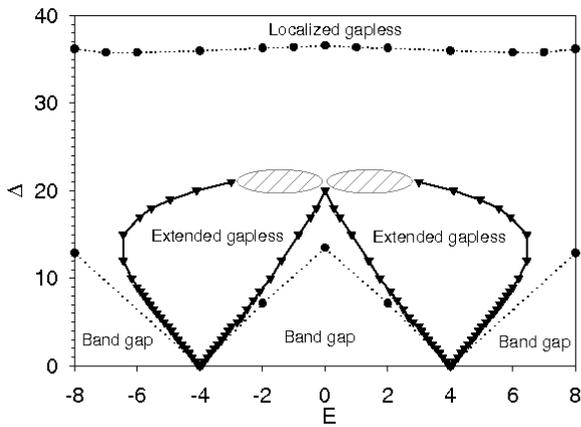}
\caption{Ground-state phase diagram as a function of energy for
$U=0$. The circles were taken from Ref. [1]. The triangles are
determined from the numerical solution of the DMFT equations for
$q=0$. The dashed region is the energy area in which we cannot
estimate the transition points accurately. Lines are guides to the
eye.} \label{phase0}
\end{figure}

\section{Results}\label{results}
First, we investigate the case without disorder. In this limit,
where $\Delta=0$, we find, independently of $q$, that the analytical
expression for the density of states of the flatband is
\begin{equation}
P_{q} (E) = \frac{E} {\pi \sqrt{[E^2 - (E_{0}+U/2)^2]
[(E_{0}-U/2)^2-E^2]} } ,
 \label{n0}
\end{equation}
if $|(E_{0}-U/2)| < |E| < |(E_{0}+U/2)|$ and $P_{q} (E)=0$,
otherwise. Typical results are shown in Fig. \ref{var1}. In order to
compare our results with Refs. [1] and [2], we have set the highly
degenerated state at $E_{0}=4$. Below $U=8$ we have two bands of
bandwidth $U$. For $U
> 8$ we have also two bands, but now the bandwidth is constant and
equal to $8$. The bandgap is always $|8-U|$. This bandgap for $U \gg
1$ exhibits the Mott insulator relationship between the bandgap and
the repulsive Coulomb potential (bandgap $\sim U$). Note that for
$U=8$ the system presents only one band of bandwidth $8$, and only
for this case $P_{q} (0)\neq 0 $.

Next, let us explore the case $\Delta \ne 0$. Here, the results are
obtained numerically. We considered as initial $P_{q}(E)$ a uniform
distribution with bandwidth greater than the Lifshitz band edge.
Then we determined $G(E)$ in order to obtain $\eta(E)$ and finally
the new values of $P_{q}(E)$. This procedure is repeated until we
find the stable configuration.

Fig. \ref{spec} shows the energy dependence of the averaged LDOS for
$U=0$ and typical values of $\Delta$. Note that the inclusion of the
disorder (i.e., $\Delta \neq 0$) suppresses the highly degenerated
localized states. We obtain the Anderson localization for a fixed
$q$ varying the disorder strength $\Delta$ for each value of $U$,
and then determining the values of $\Delta$ when $P_{q}(E)=0$. For
the highly degenerated energy state, the arithmetic mean($q=1$) of
the LDOS does not vanish at a finite critical disorder strength.
Hence, we consider it to be non-critical at the Anderson transition.
Using the geometric average ($q=0$), the LDOS vanishes at the highly
degenerated energy state for a finite value of $\Delta$. The
detection of the Anderson localization is obtained using $q=0$.

Fig. \ref{phase0} presents the phase diagram of the ground state for
the Anderson-Falikov-Kimball model  as a function of energy for
$U=0$ using the geometrical average. Our results are represented by
triangles. The results of Ref. [1] are marked by circles. As we use
an iterative process, our $P_{q}(E)$ not always converges to a
stable value for large $\Delta$. Within the dashed area of Fig. 3 we
cannot estimate the transition points accurately. We can observe
three phases: extended gapless, localized gapless and band
gap.\cite{pap2}

The DMFT gives different results from the ones obtained in Refs. [1]
and [2]. The DMFT approach reduces the extended phase, increasing
the critical disorder $\Delta_{c}$ for the
localization-delocalization transition and decreasing the one for
the delocalization-localization transition. The considerable
quantitative disagreement between both approaches requires new
investigations to better understand these differences.

\begin{figure}
\includegraphics[width=80mm]{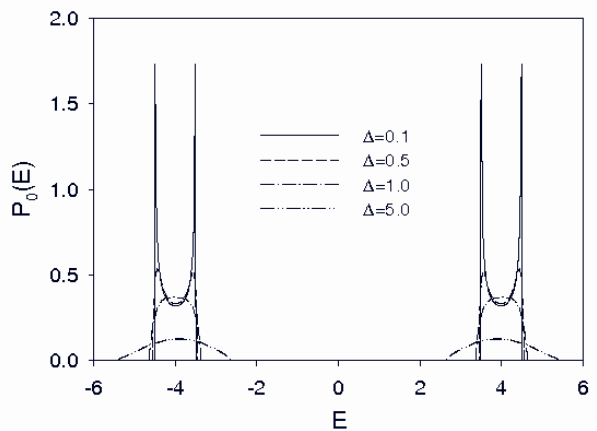}
\caption{Averaged local density of states at $U=1$ and disorder
strength $\Delta =0.1$, $0.5$, $1.0$ and $5.0$ for $q=0$.}
\label{u1}
\end{figure}

\begin{figure}
\includegraphics[width=80mm]{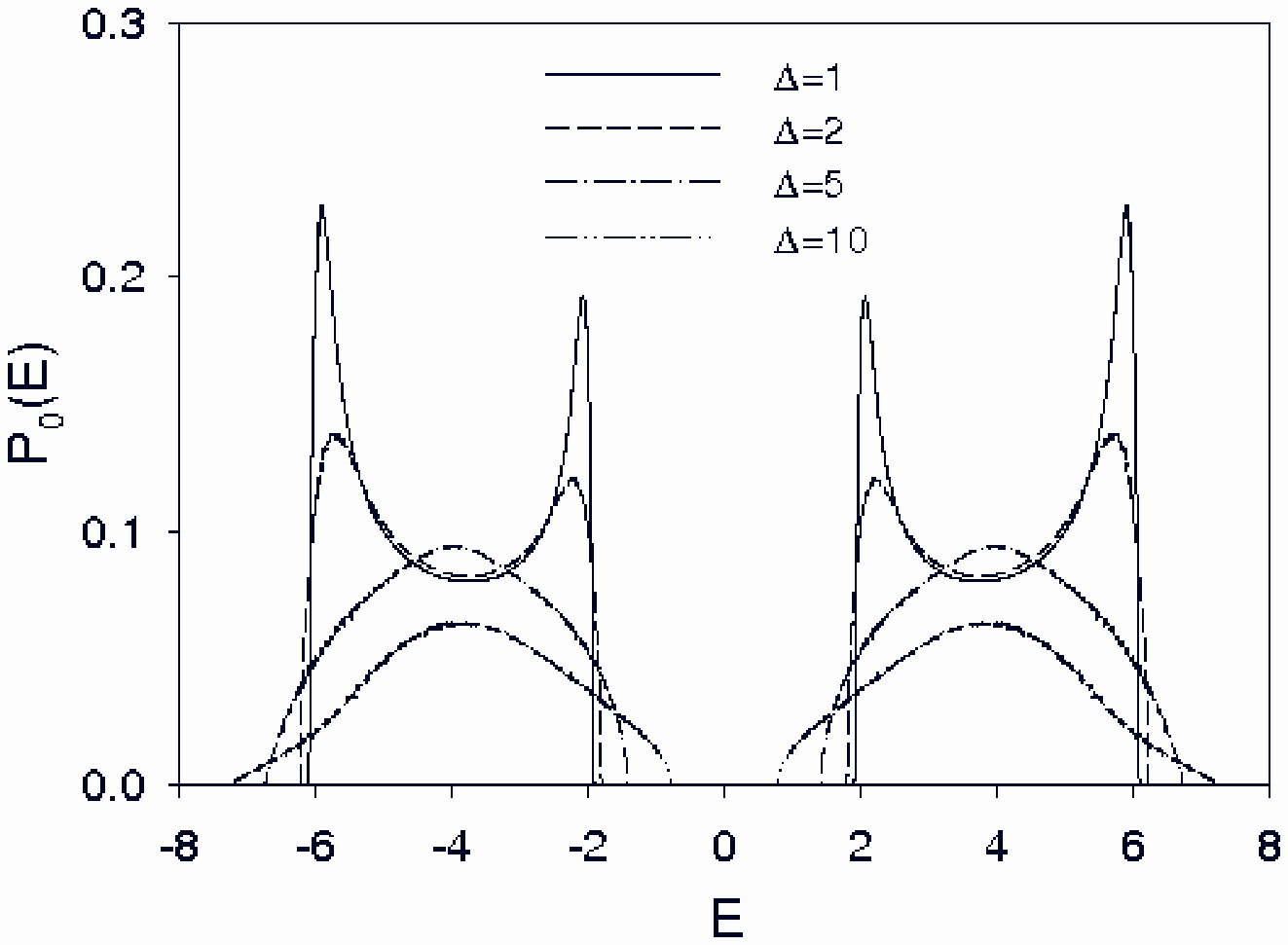}
\caption{Averaged local density of states at $U=4$ and disorder
strength $\Delta =1$, $2$, $5$ and $10$ for $q=0$.} \label{u4}
\end{figure}

\begin{figure}
\includegraphics[width=80mm]{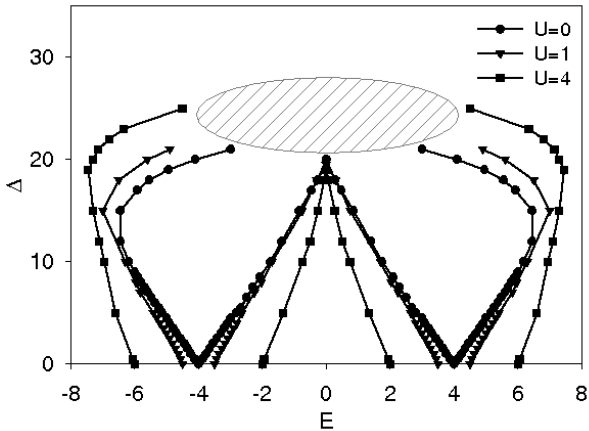}
\caption{Ground-state phase diagram as function of energy for $U=0$,
$1$ and $4$ obtained from the numerical solution of the DMFT
equations. The dashed region is the energy area in which we cannot
estimate the transition points accurately. Lines are guides to the
eye.}\label{ult}
\end{figure}

We now consider the influence of the Coulomb repulsion $U$ on the
results. Fig. \ref{u1} shows the averaged LDOS for $\Delta=0.1$,
$0.5$, $1.0$ and $5.0$ at $U=1$. The results for $U=4$ are exhibited
in Fig. \ref{u4}, for $\Delta=1$, $2$, $5$ and $10$. As already
observed for $U=0$, we find two symmetric bands. However, for small
$\Delta$, the LDOS at the end bands, corresponding to the smallest
and largest energies, are higher than those at the band centers. The
band centers correspond to the highly degenerated states for $U=0$.
Large $\Delta$ destroy the influence of the flatband and favour
higher LDOS values at band centers and smaller ones at the edge
bands. When $\Delta$ is large enough, the LDOS vanish and the
extended phase disappears. For fixed $\Delta$, the bandwidth grows
with increasing $U$.

Finally, in Figure \ref{ult} we present a complete ground-state
phase diagram for three different values of $U$, namely $U=0$, $1$
and $4$. With $U$ increases the extended phase, showing that the
interaction may enhance delocalization.

\section{Conclusions}\label{conclusions}

In the present paper, we studied the solutions of the
Anderson-Falicov-Kimball model involving a highly degenerated
localized states forming a flatband. We have shown that the new
disorder-induced insulator-metal transition of one-electron states,
called ''inverse Anderson transition'' can be obtained within
dynamical mean field theory. However, we observed a considerable
quantitative disagreement between the present approach and the
previous results.\cite{inverse,inverse2} The DMFT reduces the
extended phase, increasing the critical disorder $\Delta_{c}$ for
the localization-delocalization transition and decreasing
$\Delta_{c}$ for the delocalization-localization transition.

We also studied the dependence of the MIT transition on the
interaction between particles of the model and showed that
increasing the interaction parameter reduces the extended phase for
$U=0$.

As the doping of an impurity in the flatband has the same
characteristics of the impurity states in the quantum Hall
system\cite{inverse2}, it would be interesting to do similar
calculations for such a system.

\section{Acknowledgments}
Financial support of Conselho Nacional de Pesquisas Cientificas
(CNPq) is acknowledged.

\end{document}